\documentclass[usetimes,letter,center]{rsauthor}
\usepackage{graphicx,mathptm}
\usepackage[compress]{natbib}\bibsep=0pt
\jname{Phil.\ Tr. A}
\newcommand{\apj}{Astrophys.\ J.}
\let\apjl=\apj
\newcommand{\apjs}{Astrophys.\ J.\ Suppl.\ Ser.}
\newcommand{\aj}{Astron. J.}
\newcommand{\mnras}{Mon.\ Not.\ R.\ Astron.\ Soc.}
\newcommand{\aap}{Astron.\ Astrophys.}
\newcommand{\araa}{Ann.\ Rev.\ Astron.\ Astrophys.}
\newcommand{\nat}{Nature}
\makeatother
\begin{document}
\title{Merging White Dwarfs and Thermonuclear Supernovae}

\author{M. H. van Kerkwijk}

\address{Department of Astronomy \& Astrophysics, University of
  Toronto, 50 Saint George Street, Toronto, ON, M5S 3H4, Canada;
  mhvk@astro.utoronto.ca} 

\date{Submitted 1 June 2012}

\abstract{Thermonuclear supernovae result when interaction with a
  companion reignites nuclear fusion in a carbon-oxygen white dwarf,
  causing a thermonuclear runaway, a catastrophic gain in pressure,
  and the disintegration of the whole white dwarf.  It is usually
  thought that fusion is reignited in near-pycnonuclear conditions
  when the white dwarf approaches the Chandrasekhar mass.  I briefly
  describe two long-standing problems faced by this scenario, and our
  suggestion that these supernovae instead result from mergers of
  carbon-oxygen white dwarfs, including those that produce
  sub-Chandrasekhar mass remnants.  I then turn to possible
  observational tests, in particular those that test the absence or
  presence of electron captures during the burning.}



\maketitle

\section{The Current Paradigm and Its Problems}

A thermonuclear or type Ia supernova (SN~Ia) is generally thought to
be produced by a carbon-oxygen white dwarf that accretes matter
relatively slowly, on timescales of $\gtrsim\!10^6{\rm\,yr}$ (limited
by the rate at which heat from accretion and possible nuclear
processing can be radiated, viz., the Eddington luminosity; for
reviews, \citealt{nomo+94,hilln00}).  As the white dwarf accretes, its
interior is heated, but it does not reach ignition, because at
temperatures of $\gtrsim\!10^8{\rm\,K}$, neutrino cooling becomes
efficient enough to balance the heating (see Fig.~\ref{fig:rhot}).
However, as the white dwarf approaches the Chandrasekhar mass, the
density in its core becomes so high that fusion becomes possible at
lower temperatures (in partly pycno-nuclear conditions;
Fig.~\ref{fig:rhot}). Once this happens, a runaway ensues, stopping
only when degeneracy is lifted and thermal pressure can expand and
cool the region.  The process triggers a burning front that proceeds
through the white dwarf, generating the energy that eventually
disrupts it.

The above is physically plausible, but it has two well-known
problems.  I briefly describe these below, before turning to our alternative.

\subsection{The Paucity of Possible Progenitor Systems}

Over the age of the Universe, for every solar mass of stars formed,
$\sim\!0.0023\pm0.0006$ SN~Ia seem to occur \citep{mann+05,maoz+10}.
Since $\sim\!0.22$ white dwarfs are expected for every solar mass
formed (the remainder being in low-mass stars that are still alive),
one infers that a surprisingly high fraction, of $\sim\!1$\%, of all
white dwarfs eventually produce SN~Ia.  Comparing different galaxies,
the instantaneous SN Ia rate similarly seems to be $\sim\!1$\% of the
white-dwarf formation rate \citep{priths08}.

Most SN~Ia models invoke ``single degenerate'' progenitors, in which a
white dwarf accretes from a non-degenerate companion \citep{wheli73}.
In principle, ample numbers of such binaries exist and several routes
to explosions have been proposed \citep{ibent84}.  No route, however,
seems both common and efficient.

The main problem is that if mass transfer is slow, unstable hydrogen
fusion in the accreting matter causes novae, which in most cases
appear to remove as much mass as was accreted
(\citealt{townb04}; though white dwarfs in cataclysmic
variables are more massive than in their progenitors,
\citealt{zoro+11}).  If accretion is faster, hydrogen burns stably,
but only in a small range of accretion rate can expansion and mass
loss be avoided (\citealt{nomo+07}; for the effect of helium
flashes, see \citealt{idan+12}). Empirically, the best-suited systems
are the supersoft X-ray sources \citep{rappdss94}, but those are far
too rare to explain the SN~Ia rates \citep{dist10,gilfb10}.  We may be
missing systems, e.g., more rapidly accreting white dwarfs that
expanded and hid from X-ray view \citep{hach+10}.  However, for such
sources -- as for many single-degenerate channels -- the lack of
evidence for (entrained) hydrogen in SN~Ia is surprising (unless the
explosion can somehow be delayed, as in the ``spin-up/down'' model;
\citealt{just11,dist+11}).

Another class of SN~Ia models invoke ``double degenerates,'' where a
white dwarf merges with another \citep{webb84,ibent84}. As ignition is
not expected during the merger (except perhaps for unusually massive,
$\gtrsim\!0.9\,M_\odot$ white dwarfs, \citealt{pakm+12}), it is
usually assumed an explosion will follow only if the combined mass
exceeds the Chandrasekhar mass.  This is rare, however, and both
theoretical \citep{ruitbf09,menn+10,vkercj10} and empirical
\citep{badem12} rate estimates fall well below the SN~Ia rate.
Furthermore, the observed number of supersoft symbiotic progenitors
with the required massive white dwarfs is substantially smaller than
that expected \citep{dist10b}.

\subsection{The Difficulty of Reproducing SN~Ia Properties}

In degenerate matter, a thermonuclear runaway will proceed to
completion unless degeneracy is lifted, and thermal pressure can
expand and cool matter.  After initial ignition, what happens depends
on the conditions.  For sufficiently high overpressure in a
sufficiently large region (where what is ``sufficient'' remains to be
understood; \citealt{seit+09}), a detonation is triggered: a shock
strong enough to cause neighbouring matter to ignite and burn in turn.
Since a detonation proceeds supersonically, the white dwarf has no
time to expand and the initial density everywhere determines the
end-point of the runaway.  For a near-Chandrasekhar mass white dwarf,
most matter is at very high density and thus far too much ${}^{56}$Ni
is produced.

For a near-Chandrasekhar mass white dwarf, however, the energy release
even from fusion up to $^{56}$Ni does not lead to strong overpressure,
and a deflagration is more likely, where neighbouring regions are
ignited by a heat wave rather than a shock.  Since a deflagration is
sub-sonic, the white dwarf expands as the burning front progresses.
Thus, burning takes place at lower density, reaching lower peak
temperatures and producing less $^{56}$Ni.  Unfortunately, the burning
front appears to be too slow, making it impossible to produce
sufficiently energetic explosions \citep{hilln00}.

Another problem is that both pure detonation and pure deflagration
models do not naturally reproduce the range in SN~Ia properties, which
trace a (nearly) single-parameter family, reflecting a roughly
factor~5 range in the amount of ${}^{56}$Ni that is synthesized
\citep{phil93,mazz+07}.  The above issues can be resolved with an {\em
  ad hoc} assumption, that an initial deflagration transitions into a
detonation \citep{khok91}. If so, the timing of the transition could
determine how far the white dwarf expanded and thus how much
${}^{56}$Ni was produced.  Even with this assumption, however, it
remains unclear why the outcome would depend on the population which the
progenitor is in, i.e., why, as is observed, more luminous SN~Ia
preferentially occur in younger populations \citep{hamu+95,sull+10}.

\begin{figure}[t!]
\centerline{\includegraphics[width=0.95\hsize]{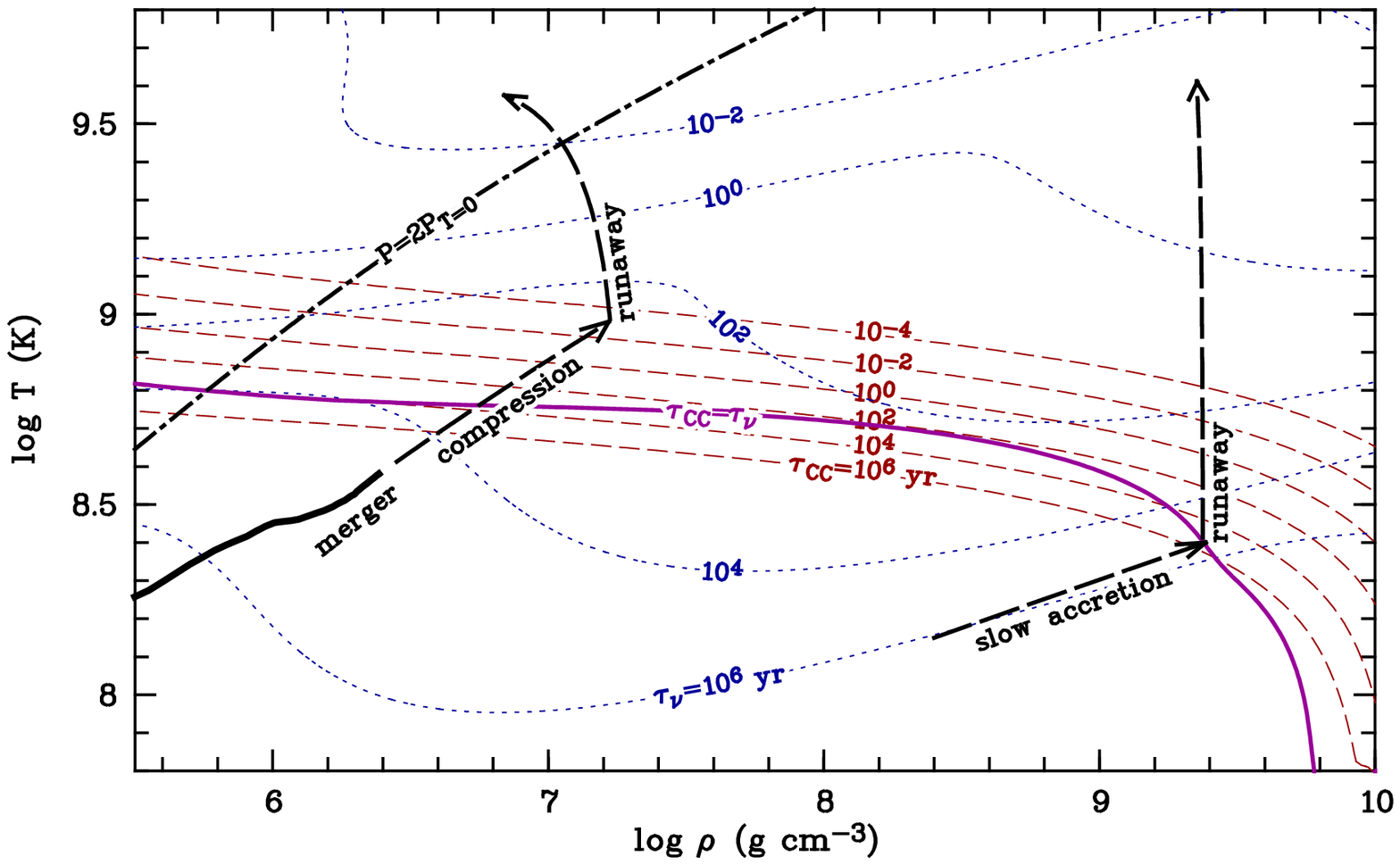}}
\caption[]{Temperate-density tracks leading to thermonuclear runaways.
  In the standard picture (on the right), a carbon-oxygen white dwarf
  accretes slowly, on a $\gtrsim\!10^6{\rm\,yr}$ timescale, and
  neutrino cooling keeps its internal temperature below a few
  $10^8$\,K (see the contours of constant cooling time $\tau_\nu$).
  When it approaches the Chandrasekhar limit and its central density
  rises dramatically, carbon fusion is ignited (where
  $\tau_{\rm{}CC}=\tau_\nu$, i.e., the fusion heating time matches the
  neutrino cooling time), and -- after 100--1000\,yr of simmering -- a
  thermal runaway ensues (at roughly constant pressure).  In our
  alternative picture \citep{vkercj10}, a white dwarf merger leads to
  a rapidly rotating remnant with a temperature-density profile like
  that shown on the left (from \citealt{loreig09}, for a merger of two
  $0.6\,M_\odot$ white dwarfs).  This is initially not hot enough for
  ignition, but as the remnant disk accretes or the core spins down,
  the interior will be compressed and heated roughly adiabatically,
  until carbon fusion becomes faster than the accretion or spin-down
  timescales and the thermonuclear runaway starts (along a curved
  constant-pressure contour, as degeneracy is lifted).
  \label{fig:rhot}}
\end{figure}

\section{Sub-Chandrasekhar Mass Mergers as SN Ia Progenitors?}

SN~Ia could be understood more easily if they arose from
sub-Chandrasekhar white dwarfs.  Since for increasing mass, a larger
fraction is dense enough to produce $^{56}$Ni
($\rho\gtrsim10^7{\rm\,g\,cm^{-3}}$), a range of $^{56}$Ni mass would
be expected.  Also, since more massive white dwarfs are the progeny of
shorter-lived stars, younger populations should preferentially host
luminous SN~Ia.  Encouragingly, pure detonations of white dwarfs with
masses between 0.9 and $1.2\,M_\odot$ reproduce the range in SN~Ia
properties, including, roughly, their lightcurves and spectra
\citep{shig+92,sim+10}.  Not clear yet, however, is whether the
distribution in luminosity can also be matched easily.

The difficulty for sub-Chandrasekhar white dwarfs is to get them hot
enough to ignite.  To overcome neutrino losses, they have to be heated
on a rather fast, $\lesssim\!10^4{\rm\,yr}$ timescale (see
Fig.~\ref{fig:rhot}).  One possibility is that carbon fusion is not
triggered directly, but indirectly, by a detonation wave started by a
thermonuclear runaway in a thick helium layer surrounding the core
\citep{woosw94}.  These ``double detonation'' models, however, predict
abundances in the outer ejecta -- produced in the helium envelope --
that are not seen in SN~Ia (\citealt{hilln00}; discussions continue
about whether these effects can be reduced by helium layers that are
thinner [\citealt{fink+10,woosk10}] or have mixed in carbon
[\citealt{krom+10}]).  Another possibility is that fusion gets ignited
during a merger that involves at least one massive,
$\gtrsim\!1\,M_\odot$ white dwarf \citep{pakm+12}.  Those, however,
have expected rates even lower than those of super-Chandrasekhar
mergers, and thus likely are too rare.

Our alternative is that SN~Ia result generally from mergers of
carbon-oxygen white dwarfs, including those with sub-Chandrasekhar
total mass \citep{vkercj10}.  Both theoretical ({\em ibid.}) and
empirical \citep{badem12} rates are a factor three or so higher than
near-Chandrasekhar rates, making them consistent with the SN~Ia rate.
Furthermore, the expected range in mass matches that for which
detonations yield sufficient ${}^{56}$Ni.  The questions are whether
fusion is ignited, and whether this triggers a detonation.

From simulations, the outcome of a white-dwarf merger depends strongly
on whether the masses are similar (where ``similar'' is within
$\sim\!0.1\,M_\odot$, \citealt{zhu+11}; Zhu {\em et al.}, 2012, in
preparation).  If they are not, the remnant consists of an almost
unaffected core of the more massive white dwarf, surrounded by a hot
envelope of the disrupted lower-mass one.  For these, further
evolution likely leads to ignition at low density, stable burning,
and, therefore, not to a SN Ia (see \citealt{shen+12}).

For similar-mass white dwarfs, however, the remnants are hot
throughout, and consist of rapidly rotating cores surrounded by thick,
dense disks.  Initially, the core is not hot enough to ignite fusion
-- nor dense enough to produce ${}^{56}$Ni -- but as the disk accretes
or the remnants spins down (helped by, e.g., strong magnetic fields
that could be generated in the strongly differentially rotating
remnant), it will be compressed and heated further (see
Fig.~\ref{fig:rhot}).  The timescale would likely be the viscous one
-- hours to days -- much faster than any relevant cooling timescale.
An open question is where ignition takes place.  If magnetic braking
is important (as in a protostar or accreting pulsar), dissipation will
be far from the remnant and ignition likely in the core.  If accretion
dominates, dissipative heating may lead to ignition in the outer
regions \citep{shen+12}.

\section{Observational Tests}

It seems unlikely that the question of the nature of the progenitors
of SN~Ia will be resolved theoretically, and hence one has to turn to
observational tests.  So far, most have focussed on trying to
distinguish between the single and double-degenerate scenario, with
conflicting results: no signature of a (former) companion in early
SN~Ia lightcurves \citep{hayd+10,bian+11,brow+12,bloo+12} or in SN Ia
remnants (e.g., \citealt{schap12,kerz+12}), yet evidence for
circumstellar medium \citep{pata+07,ster+11}.

A different test would be to distinguish between a near or
sub-Chandrasekhar mass.  One clue is that in the near-Chandrasekhar
case, where the explosion has to start with a deflagration, electron
captures during this relatively slow phase are important, leading to
the production of $\sim\!0.1\,M_\odot$ of stable iron-peak elements,
much of which is $^{58}$Ni (\citealt{maed+10b}, and references
therein).  In contrast, for sub-Chandrasekhar models, where the
density is much lower and the explosion has to be a fast detonation,
the only source of the neutrons required to produce stable iron-peak
elements is $^{22}$Ne.  This is produced during helium burning (via
${}^{14}$N$(\alpha,\gamma){}^{18}$F$({\rm e}^-\!,\nu_{\rm
  e}){}^{18}$O$(\alpha,\gamma){}^{22}$Ne, where the $^{14}$N is left
by the CNO cycle), and wherever the temperatures become hot enough to
produce $^{56}$Ni, the excess neutrons end up mostly in $^{54}$Fe and
$^{58}$Ni \citep{shig+92}, with a mass of $\sim\!(58/14)X_{\rm
  CNO}\simeq 4\%$ of the mass of $^{56}$Ni; hence, the mass of
$^{58}$Ni should be $\lesssim\!0.02\,M_\odot$ for a typical SN~Ia with
$0.6\,M_\odot$ of $^{56}$Ni.

Given the above, an observational test would be to look for evidence
for a core dominated by stable elements.  Arguably the most direct
measurement of the amount of $^{58}$Ni has been done from mid-infrared
fine-structure lines in SN 2005df (in the nebular phase, when all
$^{56}$Ni has decayed; note that in other analyses often a
near-Chandrasekhar explosion is assumed indirectly, e.g., in using the
W7 model [e.g., \citealt{mazz+07}]).  These yield an estimate of
$\sim\!0.01\,M_\odot$ of nickel, which is much more consistent with a
sub-Chandrasekhar model (\citealt{gera+07}; note that these authors
argued even this small mass was evidence for electron captures, but
they did not consider the effect of $^{22}$Ne).  Similarly, the
meteoritic abundance of nickel is $\sim\!5$\% that of iron
\citep{cox+00}, which is more easily understood in sub-Chandrasekhar
models (as already noted in, e.g., \citealt{shig+92,nomo+94}).

In contrast, the presence of an inert, colder core is inferred from
flat-topped line profiles \citep{moto+06}.  It is unclear, however,
whether this cold core reflects a lack of heating, or rather enhanced
cooling in an ``infrared catastrophe'' \citep{lelo+09}.  Evidence for
an inert core comes also from differences in line profiles for lower
and higher ionisation states \citep{maed+10a}, differences that
correlate with other SN Ia properties and are plausible for delayed
detonation, near-Chandrasekhar models \citep{maed+10c}.  It is not yet
known what to expect for sub-Chandrasekhar explosions, but nebular
spectroscopy nevertheless seems one of the most promising ways of
determining whether SN~Ia result from near or sub-Chandrasekhar mass
objects.  Ideally, one would study supernovae that cover not only a
range in SN~Ia properties but also in host metallicity (with which
$^{58}$Ni should scale linearly for sub-Chandrasekhar models; for
near-Chandrasekhar models, the dependence is more complicated, see,
e.g., \citealt{jack+10}).

\ack{I thank Ken Nomoto and Kei'ichi Maeda for filling in many lacunae
  in my understanding of SN Ia and for pointing out to the importance
  of electron captures, Michael Lennox for help in researching the
  literature, Charles Zhu, Wolfgang Kerzendorf and Stuart Sim for
  discussions, and the referees for useful comments.}


\begin{thebibliography}{55}
\providecommand{\natexlab}[1]{#1}
\expandafter\ifx\csname urlstyle\endcsname\relax
  \providecommand{\doi}[1]{doi:\discretionary{}{}{}#1}\else
  \providecommand{\doi}{doi:\discretionary{}{}{}\begingroup
  \urlstyle{rm}\Url}\fi

\bibitem[{{Badenes} \& {Maoz}(2012)}]{badem12}
{Badenes}, C. \& {Maoz}, D. 2012 {The Merger Rate of Binary White Dwarfs in the
  Galactic Disk}.
\newblock \emph{\apjl}, \textbf{749}, L11.
\newblock (\doi{10.1088/2041-8205/749/1/L11})

\bibitem[{{Bianco} \emph{et~al.}(2011)}]{bian+11}
{Bianco}, F.~B. \emph{et~al.} 2011 {Constraining Type Ia Supernovae Progenitors
  from Three Years of Supernova Legacy Survey Data}.
\newblock \emph{\apj}, \textbf{741}, 20.
\newblock (\doi{10.1088/0004-637X/741/1/20})

\bibitem[{{Bloom} \emph{et~al.}(2012)}]{bloo+12}
{Bloom}, J.~S. \emph{et~al.} 2012 {A Compact Degenerate Primary-star Progenitor
  of SN 2011fe}.
\newblock \emph{\apjl}, \textbf{744}, L17.
\newblock (\doi{10.1088/2041-8205/744/2/L17})

\bibitem[{{Brown} \emph{et~al.}(2012){Brown}, {Dawson}, {Harris}, {Olmstead},
  {Milne} \& {Roming}}]{brow+12}
{Brown}, P.~J., {Dawson}, K.~S., {Harris}, D.~W., {Olmstead}, M., {Milne}, P.
  \& {Roming}, P.~W.~A. 2012 {Constraints on Type Ia Supernova Progenitor
  Companions from Early Ultraviolet Observations with Swift}.
\newblock \emph{\apj}, \textbf{749}, 18.
\newblock (\doi{10.1088/0004-637X/749/1/18})

\bibitem[{{Cox}(2000)}]{cox+00}
{Cox}, A.~N. 2000 \emph{{Allen's astrophysical quantities}}.

\bibitem[{{Di Stefano}(2010{\natexlab{\emph{a}}})}]{dist10}
{Di Stefano}, R. 2010{\natexlab{\emph{a}}} {The Progenitors of Type Ia
  Supernovae. I. Are they Supersoft Sources?}
\newblock \emph{\apj}, \textbf{712}, 728--733.
\newblock (\doi{10.1088/0004-637X/712/1/728})

\bibitem[{{Di Stefano}(2010{\natexlab{\emph{b}}})}]{dist10b}
{Di Stefano}, R. 2010{\natexlab{\emph{b}}} {The Progenitors of Type Ia
  Supernovae. II. Are they Double-degenerate Binaries? The Symbiotic Channel}.
\newblock \emph{\apj}, \textbf{719}, 474--482.
\newblock (\doi{10.1088/0004-637X/719/1/474})

\bibitem[{{Di Stefano} \emph{et~al.}(2011){Di Stefano}, {Voss} \&
  {Claeys}}]{dist+11}
{Di Stefano}, R., {Voss}, R. \& {Claeys}, J.~S.~W. 2011 {Spin-up/Spin-down
  Models for Type Ia Supernovae}.
\newblock \emph{\apjl}, \textbf{738}, L1.
\newblock (\doi{10.1088/2041-8205/738/1/L1})

\bibitem[{{Fink} \emph{et~al.}(2010){Fink}, {R{\"o}pke}, {Hillebrandt},
  {Seitenzahl}, {Sim} \& {Kromer}}]{fink+10}
{Fink}, M., {R{\"o}pke}, F.~K., {Hillebrandt}, W., {Seitenzahl}, I.~R., {Sim},
  S.~A. \& {Kromer}, M. 2010 {Double-detonation sub-Chandrasekhar supernovae:
  can minimum helium shell masses detonate the core?}
\newblock \emph{\aap}, \textbf{514}, A53.
\newblock (\doi{10.1051/0004-6361/200913892})

\bibitem[{{Gerardy} \emph{et~al.}(2007)}]{gera+07}
{Gerardy}, C.~L. \emph{et~al.} 2007 {Signatures of Delayed Detonation,
  Asymmetry, and Electron Capture in the Mid-Infrared Spectra of Supernovae
  2003hv and 2005df}.
\newblock \emph{\apj}, \textbf{661}, 995--1012.
\newblock (\doi{10.1086/516728})

\bibitem[{{Gilfanov} \& {Bogd{\'a}n}(2010)}]{gilfb10}
{Gilfanov}, M. \& {Bogd{\'a}n}, {\'A}. 2010 {An upper limit on the contribution
  of accreting white dwarfs to the typeIa supernova rate}.
\newblock \emph{\nat}, \textbf{463}, 924--925.
\newblock (\doi{10.1038/nature08685})

\bibitem[{{Hachisu} \emph{et~al.}(2010){Hachisu}, {Kato} \& {Nomoto}}]{hach+10}
{Hachisu}, I., {Kato}, M. \& {Nomoto}, K. 2010 {Supersoft X-ray Phase of Single
  Degenerate Type Ia Supernova Progenitors in Early-type Galaxies}.
\newblock \emph{\apjl}, \textbf{724}, L212--L216.
\newblock (\doi{10.1088/2041-8205/724/2/L212})

\bibitem[{{Hamuy} \emph{et~al.}(1995){Hamuy}, {Phillips}, {Maza}, {Suntzeff},
  {Schommer} \& {Aviles}}]{hamu+95}
{Hamuy}, M., {Phillips}, M.~M., {Maza}, J., {Suntzeff}, N.~B., {Schommer},
  R.~A. \& {Aviles}, R. 1995 {A Hubble diagram of distant type IA supernovae}.
\newblock \emph{\aj}, \textbf{109}, 1--13.
\newblock (\doi{10.1086/117251})

\bibitem[{{Hayden} \emph{et~al.}(2010)}]{hayd+10}
{Hayden}, B.~T. \emph{et~al.} 2010 {Single or Double Degenerate Progenitors?
  Searching for Shock Emission in the SDSS-II Type Ia Supernovae}.
\newblock \emph{\apj}, \textbf{722}, 1691--1698.
\newblock (\doi{10.1088/0004-637X/722/2/1691})

\bibitem[{{Hillebrandt} \& {Niemeyer}(2000)}]{hilln00}
{Hillebrandt}, W. \& {Niemeyer}, J.~C. 2000 {Type IA Supernova Explosion
  Models}.
\newblock \emph{\araa}, \textbf{38}, 191--230.
\newblock (\doi{10.1146/annurev.astro.38.1.191})

\bibitem[{{Iben}, Jr. \& {Tutukov}(1984){Iben} \& {Tutukov}}]{ibent84}
{Iben}, Jr., I. \& {Tutukov}, A.~V. 1984 {Supernovae of type I as end products
  of the evolution of binaries with components of moderate initial mass (M not
  greater than about 9 solar masses)}.
\newblock \emph{\apjs}, \textbf{54}, 335--372.
\newblock (\doi{10.1086/190932})

\bibitem[{{Idan} \emph{et~al.}(2012){Idan}, {Shaviv} \& {Shaviv}}]{idan+12}
{Idan}, I., {Shaviv}, N.~J. \& {Shaviv}, G. 2012 {The Fate of a WD Accreting
  H-Rich Material at High Rates}.
\newblock \emph{Journal of Physics Conference Series}, \textbf{337}(1),
  012\,051.
\newblock (\doi{10.1088/1742-6596/337/1/012051})

\bibitem[{{Jackson} \emph{et~al.}(2010){Jackson}, {Calder}, {Townsley},
  {Chamulak}, {Brown} \& {Timmes}}]{jack+10}
{Jackson}, A.~P., {Calder}, A.~C., {Townsley}, D.~M., {Chamulak}, D.~A.,
  {Brown}, E.~F. \& {Timmes}, F.~X. 2010 {Evaluating Systematic Dependencies of
  Type Ia Supernovae: The Influence of Deflagration to Detonation Density}.
\newblock \emph{\apj}, \textbf{720}, 99--113.
\newblock (\doi{10.1088/0004-637X/720/1/99})

\bibitem[{{Justham}(2011)}]{just11}
{Justham}, S. 2011 {Single-degenerate Type Ia Supernovae Without Hydrogen
  Contamination}.
\newblock \emph{\apjl}, \textbf{730}, L34.
\newblock (\doi{10.1088/2041-8205/730/2/L34})

\bibitem[{{Kerzendorf} \emph{et~al.}(2012){Kerzendorf}, {Schmidt}, {Laird},
  {Podsiadlowski} \& {Bessell}}]{kerz+12}
{Kerzendorf}, W.~E., {Schmidt}, B.~P., {Laird}, J.~B., {Podsiadlowski}, P. \&
  {Bessell}, M.~S. 2012 {Hunting for the progenitor of SN 1006: High resolution
  spectroscopic search with the FLAMES instrument}.
\newblock \emph{arXiv:1207.4481}.

\bibitem[{{Khokhlov}(1991)}]{khok91}
{Khokhlov}, A.~M. 1991 {Delayed detonation model for type IA supernovae}.
\newblock \emph{\aap}, \textbf{245}, 114--128.

\bibitem[{{Kromer} \emph{et~al.}(2010){Kromer}, {Sim}, {Fink}, {R{\"o}pke},
  {Seitenzahl} \& {Hillebrandt}}]{krom+10}
{Kromer}, M., {Sim}, S.~A., {Fink}, M., {R{\"o}pke}, F.~K., {Seitenzahl}, I.~R.
  \& {Hillebrandt}, W. 2010 {Double-detonation Sub-Chandrasekhar Supernovae:
  Synthetic Observables for Minimum Helium Shell Mass Models}.
\newblock \emph{\apj}, \textbf{719}, 1067--1082.
\newblock (\doi{10.1088/0004-637X/719/2/1067})

\bibitem[{{Leloudas} \emph{et~al.}(2009)}]{lelo+09}
{Leloudas}, G. \emph{et~al.} 2009 {The normal Type Ia SN 2003hv out to very
  late phases}.
\newblock \emph{\aap}, \textbf{505}, 265--279.
\newblock (\doi{10.1051/0004-6361/200912364})

\bibitem[{{Lor{\'e}n-Aguilar} \emph{et~al.}(2009){Lor{\'e}n-Aguilar}, {Isern}
  \& {Garc{\'{\i}}a-Berro}}]{loreig09}
{Lor{\'e}n-Aguilar}, P., {Isern}, J. \& {Garc{\'{\i}}a-Berro}, E. 2009
  {High-resolution smoothed particle hydrodynamics simulations of the merger of
  binary white dwarfs}.
\newblock \emph{\aap}, \textbf{500}, 1193--1205.
\newblock (\doi{10.1051/0004-6361/200811060})

\bibitem[{{Maeda} \emph{et~al.}(2010{\natexlab{\emph{a}}}){Maeda}, {R{\"o}pke},
  {Fink}, {Hillebrandt}, {Travaglio} \& {Thielemann}}]{maed+10b}
{Maeda}, K., {R{\"o}pke}, F.~K., {Fink}, M., {Hillebrandt}, W., {Travaglio}, C.
  \& {Thielemann}, F.-K. 2010{\natexlab{\emph{a}}} {Nucleosynthesis in
  Two-Dimensional Delayed Detonation Models of Type Ia Supernova Explosions}.
\newblock \emph{\apj}, \textbf{712}, 624--638.
\newblock (\doi{10.1088/0004-637X/712/1/624})

\bibitem[{{Maeda} \emph{et~al.}(2010{\natexlab{\emph{b}}}){Maeda},
  {Taubenberger}, {Sollerman}, {Mazzali}, {Leloudas}, {Nomoto} \&
  {Motohara}}]{maed+10a}
{Maeda}, K., {Taubenberger}, S., {Sollerman}, J., {Mazzali}, P.~A., {Leloudas},
  G., {Nomoto}, K. \& {Motohara}, K. 2010{\natexlab{\emph{b}}} {Nebular Spectra
  and Explosion Asymmetry of Type Ia Supernovae}.
\newblock \emph{\apj}, \textbf{708}, 1703--1715.
\newblock (\doi{10.1088/0004-637X/708/2/1703})

\bibitem[{{Maeda} \emph{et~al.}(2010{\natexlab{\emph{c}}})}]{maed+10c}
{Maeda}, K. \emph{et~al.} 2010{\natexlab{\emph{c}}} {An asymmetric explosion as
  the origin of spectral evolution diversity in type Ia supernovae}.
\newblock \emph{\nat}, \textbf{466}, 82--85.
\newblock (\doi{10.1038/nature09122})

\bibitem[{{Mannucci} \emph{et~al.}(2005)}]{mann+05}
{Mannucci}, F. \emph{et~al.} 2005 {The supernova rate per unit mass}.
\newblock \emph{\aap}, \textbf{433}, 807--814.
\newblock (\doi{10.1051/0004-6361:20041411})

\bibitem[{{Maoz} \emph{et~al.}(2011){Maoz}, {Mannucci}, {Li}, {Filippenko},
  {Della Valle} \& {Panagia}}]{maoz+10}
{Maoz}, D., {Mannucci}, F., {Li}, W., {Filippenko}, A.~V., {Della Valle}, M. \&
  {Panagia}, N. 2011 {Nearby supernova rates from the Lick Observatory
  Supernova Search - IV. A recovery method for the delay-time distribution}.
\newblock \emph{\mnras}, \textbf{412}, 1508--1521.
\newblock (\doi{10.1111/j.1365-2966.2010.16808.x})

\bibitem[{{Mazzali} \emph{et~al.}(2007){Mazzali}, {R{\"o}pke}, {Benetti} \&
  {Hillebrandt}}]{mazz+07}
{Mazzali}, P.~A., {R{\"o}pke}, F.~K., {Benetti}, S. \& {Hillebrandt}, W. 2007
  {A Common Explosion Mechanism for Type Ia Supernovae}.
\newblock \emph{Science}, \textbf{315}, 825.
\newblock (\doi{10.1126/science.1136259})

\bibitem[{{Mennekens} \emph{et~al.}(2010){Mennekens}, {Vanbeveren}, {De Greve}
  \& {De Donder}}]{menn+10}
{Mennekens}, N., {Vanbeveren}, D., {De Greve}, J.~P. \& {De Donder}, E. 2010
  {The delay-time distribution of Type Ia supernovae: a comparison between
  theory and observation}.
\newblock \emph{\aap}, \textbf{515}, A89.
\newblock (\doi{10.1051/0004-6361/201014115})

\bibitem[{{Motohara} \emph{et~al.}(2006)}]{moto+06}
{Motohara}, K. \emph{et~al.} 2006 {The Asymmetric Explosion of Type Ia
  Supernovae as Seen from Near-Infrared Observations}.
\newblock \emph{\apjl}, \textbf{652}, L101--L104.
\newblock (\doi{10.1086/509919})

\bibitem[{{Nomoto} \emph{et~al.}(2007){Nomoto}, {Saio}, {Kato} \&
  {Hachisu}}]{nomo+07}
{Nomoto}, K., {Saio}, H., {Kato}, M. \& {Hachisu}, I. 2007 {Thermal Stability
  of White Dwarfs Accreting Hydrogen-rich Matter and Progenitors of Type Ia
  Supernovae}.
\newblock \emph{\apj}, \textbf{663}, 1269--1276.
\newblock (\doi{10.1086/518465})

\bibitem[{{Nomoto} \emph{et~al.}(1994){Nomoto}, {Yamaoka}, {Shigeyama},
  {Kumagai} \& {Tsujimoto}}]{nomo+94}
{Nomoto}, K., {Yamaoka}, H., {Shigeyama}, T., {Kumagai}, S. \& {Tsujimoto}, T.
  1994 {Type I supernovae and evolution of interacting binaries}.
\newblock In \emph{Supernovae} (eds S.~A. {Bludman}, R.~{Mochkovitch} \&
  J.~{Zinn-Justin}), p. 199.

\bibitem[{{Pakmor} \emph{et~al.}(2012){Pakmor}, {Kromer}, {Taubenberger},
  {Sim}, {R{\"o}pke} \& {Hillebrandt}}]{pakm+12}
{Pakmor}, R., {Kromer}, M., {Taubenberger}, S., {Sim}, S.~A., {R{\"o}pke},
  F.~K. \& {Hillebrandt}, W. 2012 {Normal Type Ia Supernovae from Violent
  Mergers of White Dwarf Binaries}.
\newblock \emph{\apjl}, \textbf{747}, L10.
\newblock (\doi{10.1088/2041-8205/747/1/L10})

\bibitem[{{Patat} \emph{et~al.}(2007)}]{pata+07}
{Patat}, F. \emph{et~al.} 2007 {Detection of Circumstellar Material in a Normal
  Type Ia Supernova}.
\newblock \emph{Science}, \textbf{317}, 924.
\newblock (\doi{10.1126/science.1143005})

\bibitem[{{Phillips}(1993)}]{phil93}
{Phillips}, M.~M. 1993 {The absolute magnitudes of Type IA supernovae}.
\newblock \emph{\apjl}, \textbf{413}, L105--L108.
\newblock (\doi{10.1086/186970})

\bibitem[{{Pritchet} \emph{et~al.}(2008){Pritchet}, {Howell} \&
  {Sullivan}}]{priths08}
{Pritchet}, C.~J., {Howell}, D.~A. \& {Sullivan}, M. 2008 {The Progenitors of
  Type Ia Supernovae}.
\newblock \emph{\apjl}, \textbf{683}, L25--L28.
\newblock (\doi{10.1086/591314})

\bibitem[{{Rappaport} \emph{et~al.}(1994){Rappaport}, {Di Stefano} \&
  {Smith}}]{rappdss94}
{Rappaport}, S., {Di Stefano}, R. \& {Smith}, J.~D. 1994 {Formation and
  evolution of luminous supersoft X-ray sources}.
\newblock \emph{\apj}, \textbf{426}, 692--703.
\newblock (\doi{10.1086/174106})

\bibitem[{{Ruiter} \emph{et~al.}(2009){Ruiter}, {Belczynski} \&
  {Fryer}}]{ruitbf09}
{Ruiter}, A.~J., {Belczynski}, K. \& {Fryer}, C. 2009 {Rates and Delay Times of
  Type Ia Supernovae}.
\newblock \emph{\apj}, \textbf{699}, 2026--2036.
\newblock (\doi{10.1088/0004-637X/699/2/2026})

\bibitem[{{Schaefer} \& {Pagnotta}(2012)}]{schap12}
{Schaefer}, B.~E. \& {Pagnotta}, A. 2012 {An absence of ex-companion stars in
  the type Ia supernova remnant SNR 0509-67.5}.
\newblock \emph{\nat}, \textbf{481}, 164--166.
\newblock (\doi{10.1038/nature10692})

\bibitem[{{Seitenzahl} \emph{et~al.}(2009){Seitenzahl}, {Meakin}, {Townsley},
  {Lamb} \& {Truran}}]{seit+09}
{Seitenzahl}, I.~R., {Meakin}, C.~A., {Townsley}, D.~M., {Lamb}, D.~Q. \&
  {Truran}, J.~W. 2009 {Spontaneous Initiation of Detonations in White Dwarf
  Environments: Determination of Critical Sizes}.
\newblock \emph{\apj}, \textbf{696}, 515--527.
\newblock (\doi{10.1088/0004-637X/696/1/515})

\bibitem[{{Shen} \emph{et~al.}(2012){Shen}, {Bildsten}, {Kasen} \&
  {Quataert}}]{shen+12}
{Shen}, K.~J., {Bildsten}, L., {Kasen}, D. \& {Quataert}, E. 2012 {The
  Long-term Evolution of Double White Dwarf Mergers}.
\newblock \emph{\apj}, \textbf{748}, 35.
\newblock (\doi{10.1088/0004-637X/748/1/35})

\bibitem[{{Shigeyama} \emph{et~al.}(1992){Shigeyama}, {Nomoto}, {Yamaoka} \&
  {Thielemann}}]{shig+92}
{Shigeyama}, T., {Nomoto}, K., {Yamaoka}, H. \& {Thielemann}, F.-K. 1992
  {Possible models for the type IA supernova 1990N}.
\newblock \emph{\apjl}, \textbf{386}, L13--L16.
\newblock (\doi{10.1086/186281})

\bibitem[{{Sim} \emph{et~al.}(2010)}]{sim+10}
{Sim}, S.~A. \emph{et~al.} 2010 {Detonations in Sub-Chandrasekhar-mass C+O
  White Dwarfs}.
\newblock \emph{\apjl}, \textbf{714}, L52--L57.
\newblock (\doi{10.1088/2041-8205/714/1/L52})

\bibitem[{{Sternberg} \emph{et~al.}(2011)}]{ster+11}
{Sternberg}, A. \emph{et~al.} 2011 {Circumstellar Material in Type Ia
  Supernovae via Sodium Absorption Features}.
\newblock \emph{Science}, \textbf{333}, 856.
\newblock (\doi{10.1126/science.1203836})

\bibitem[{{Sullivan} \emph{et~al.}(2010)}]{sull+10}
{Sullivan}, M. \emph{et~al.} 2010 {The dependence of Type Ia Supernovae
  luminosities on their host galaxies}.
\newblock \emph{\mnras}, \textbf{406}, 782--802.
\newblock (\doi{10.1111/j.1365-2966.2010.16731.x})

\bibitem[{{Townsley} \& {Bildsten}(2004)}]{townb04}
{Townsley}, D.~M. \& {Bildsten}, L. 2004 {Theoretical Modeling of the Thermal
  State of Accreting White Dwarfs Undergoing Classical Nova Cycles}.
\newblock \emph{\apj}, \textbf{600}, 390--403.
\newblock (\doi{10.1086/379647})

\bibitem[{{van Kerkwijk} \emph{et~al.}(2010){van Kerkwijk}, {Chang} \&
  {Justham}}]{vkercj10}
{van Kerkwijk}, M.~H., {Chang}, P. \& {Justham}, S. 2010 {Sub-Chandrasekhar
  White Dwarf Mergers as the Progenitors of Type Ia Supernovae}.
\newblock \emph{\apjl}, \textbf{722}, L157--L161.
\newblock (\doi{10.1088/2041-8205/722/2/L157})

\bibitem[{{Webbink}(1984)}]{webb84}
{Webbink}, R.~F. 1984 {Double white dwarfs as progenitors of R Coronae Borealis
  stars and Type I supernovae}.
\newblock \emph{\apj}, \textbf{277}, 355--360.
\newblock (\doi{10.1086/161701})

\bibitem[{{Whelan} \& {Iben}(1973)}]{wheli73}
{Whelan}, J. \& {Iben}, Jr., I. 1973 {Binaries and Supernovae of Type I}.
\newblock \emph{\apj}, \textbf{186}, 1007--1014.
\newblock (\doi{10.1086/152565})

\bibitem[{{Woosley} \& {Kasen}(2011)}]{woosk10}
{Woosley}, S.~E. \& {Kasen}, D. 2011 {Sub-Chandrasekhar Mass Models for
  Supernovae}.
\newblock \emph{\apj}, \textbf{734}, 38.
\newblock (\doi{10.1088/0004-637X/734/1/38})

\bibitem[{{Woosley} \& {Weaver}(1994)}]{woosw94}
{Woosley}, S.~E. \& {Weaver}, T.~A. 1994 {Sub-Chandrasekhar mass models for
  Type IA supernovae}.
\newblock \emph{\apj}, \textbf{423}, 371--379.
\newblock (\doi{10.1086/173813})

\bibitem[{{Zhu} \emph{et~al.}(2011){Zhu}, {Chang}, {van Kerkwijk} \&
  {Wadsley}}]{zhu+11}
{Zhu}, C., {Chang}, P., {van Kerkwijk}, M. \& {Wadsley}, J. 2011 {Properties of
  Carbon-Oxygen White Dwarf Merger Remnants}.
\newblock \emph{arXiv:1109.4334}.

\bibitem[{{Zorotovic} \emph{et~al.}(2011){Zorotovic}, {Schreiber} \&
  {G{\"a}nsicke}}]{zoro+11}
{Zorotovic}, M., {Schreiber}, M.~R. \& {G{\"a}nsicke}, B.~T. 2011 {Post common
  envelope binaries from SDSS. XI. The white dwarf mass distributions of CVs
  and pre-CVs}.
\newblock \emph{\aap}, \textbf{536}, A42.
\newblock (\doi{10.1051/0004-6361/201116626})

\end{thebibliography}

\end{document}